\documentclass[11pt,a4paper]{article}
\usepackage[utf8]{inputenc}
\usepackage{graphicx}
\usepackage{amsmath}
\usepackage{hyperref}
\usepackage{geometry}
\begin{document}

\begin{center}
{\Large \textbf{Large Language Model Integration for Knowledge Retrieval and Interaction for the DUNE Experiment}}\\[0.5cm]
\textbf{A. Rafique, A. Singh, R. Srinivas for the DUNE Collaboration}\\[0.3cm]
{\small Argonne National Laboratory}\\[0.3cm]
Presented at the 32nd International Symposium on Lepton Photon Interactions at High Energies, Madison, Wisconsin, USA, August 25--29, 2025\\[1cm]
\end{center}

\begin{abstract}
The Deep Underground Neutrino Experiment (DUNE) is a next-generation neutrino experiment that will generate an unprecedented volume of heterogeneous information—from documentation and technical notes to experimental data and reconstruction pipelines. Efficient knowledge retrieval and contextual understanding are increasingly critical for collaboration-wide productivity and onboarding. In this work, we present \textbf{DUNE-GPT}, a prototype framework that leverages large language models (LLMs) and retrieval-augmented generation (RAG) to enable natural-language querying of DUNE’s internal documentation and technical resources. The system provides an intelligent interface for DUNE collaborators to interact with experiment-specific knowledge while maintaining data privacy and infrastructure compliance within Fermilab computing resources.
\end{abstract}

\section{Introduction}
As high-energy physics (HEP) experiments grow in complexity and scale, the ability to access and interpret collaboration-specific knowledge becomes increasingly challenging. The DUNE collaboration~\cite{dune_tdr} maintains an extensive set of documents hosted across multiple platforms, including DocDB, Indico, and internal wikis. Navigating this distributed information landscape can be time-consuming, especially for new collaborators or when searching for technical details on reconstruction, simulation, data analysis, and detector operations.

\begin{figure}[h!]
    \centering
    \includegraphics[width=0.5\textwidth]{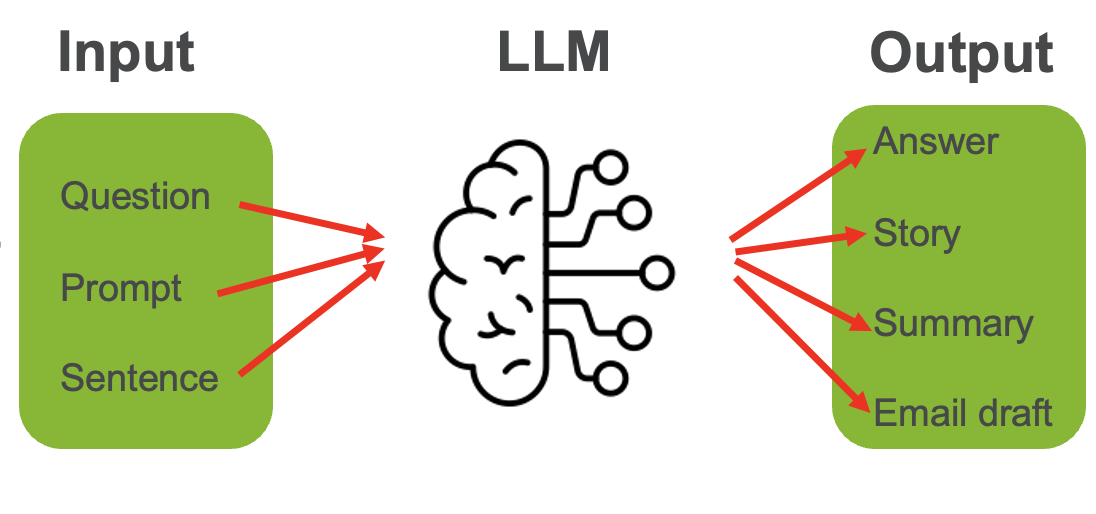}
    \caption{Example of input and output types for a Large Language Model (LLM).}
    \label{fig:LLM}
\end{figure}

Recent advances in large language models (LLMs)~\cite{openai_gpt} have demonstrated remarkable capabilities in contextual reasoning and natural-language retrieval. Figure~\ref{fig:LLM} illustrates the types of inputs and outputs an LLM can generate. However, direct deployment of commercial or open LLMs in HEP collaborations raises concerns related to data privacy, reproducibility, and network accessibility. To address these challenges, we developed \textbf{DUNE-GPT}, an internal AI assistant designed to operate within Fermilab’s computing environment, providing experiment-specific question-answering capabilities powered by retrieval-augmented generation (RAG)~\cite{rag_paper}.

\section{System Overview}
The DUNE-GPT framework consists of three main components:
\begin{enumerate}
    \item Document processing and embedding generation
    \item Vector database and retrieval engine
    \item LLM interface for response generation
\end{enumerate}

\begin{figure}[h!]
    \centering
    \includegraphics[width=0.8\textwidth]{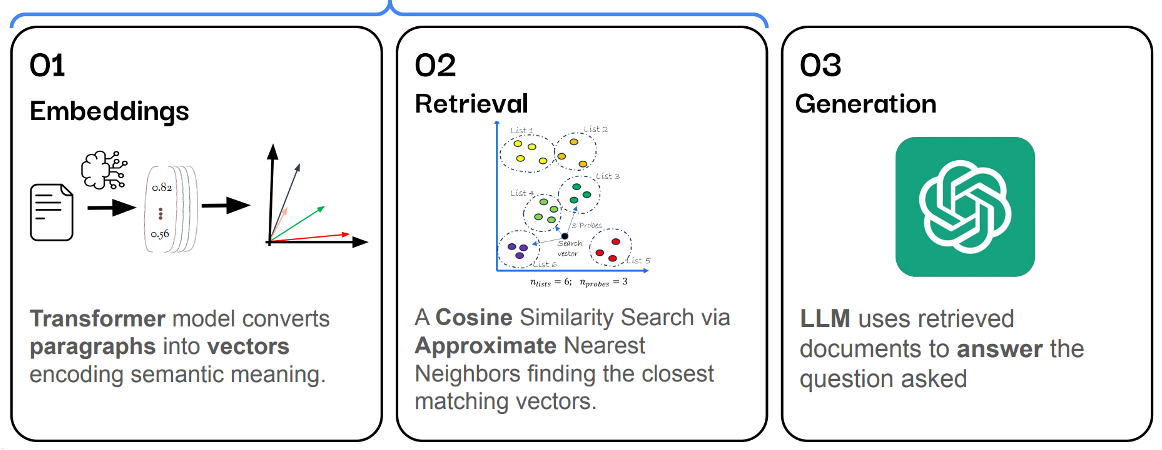}
    \caption{Overview of the Retrieval-Augmented Generation (RAG) pipeline. Embeddings are created first, followed by retrieval of the most relevant documents upon user query, which are then passed to the LLM for final response generation.This figure is reproduced from the publicly available chATLAS presentation~\cite{DalSanto:2935252}.}
    \label{fig:RAG}
\end{figure}

Figure~\ref{fig:RAG} presents these three components and gives an overview of each step of the process. 

\subsection{Data Sources}
We extracted publicly accessible and internal DUNE documentation, including DUNE documents, presentations, meeting notes, technical design reports, and working group materials from DocDB and Indico. Text preprocessing involved parsing multiple formats (PDF, DOCX, TXT, PNG, and others), metadata extraction, and token-level segmentation for embedding generation. Sensitive or restricted content was excluded in accordance with collaboration policy. Only documents accessible to the entire collaboration were processed. 

\subsection{Embedding and Retrieval Layer}
Each processed document was embedded using transformer-based encoders (multi-qa-mpnet-base-dot-v1 for this study) optimized for scientific text. These embeddings were stored in a vector database, Facebook AI Similarity Search (FAISS), to enable efficient similarity search. When a user issues a query, the system retrieves the most relevant text segments using cosine similarity and forwards them to the response generation module.

\subsection{Large Language Model Integration}
The response generation step employs pre-existing LLMs hosted on Argonne and Fermilab machines (Argo~\cite{argo} and Ollama~\cite{ollama}, respectively). The system performs RAG by conditioning the LLM on retrieved context snippets, ensuring that generated responses remain grounded in verified DUNE sources. This architecture enables controlled, explainable responses while minimizing the risk of hallucination.

\section{Implementation and Deployment}
The prototype was initially developed using the Aurora~\cite{aurora} supercomputer at the Argonne Leadership Computing Facility (ALCF)~\cite{alcf}, and will later be migrated to Fermilab infrastructure for long-term hosting. The backend is implemented in Python, integrating open-source libraries for embedding generation and retrieval. A lightweight web interface allows collaborators to issue natural-language queries and receive formatted responses that include document citations from within DUNE internal databases.

To align with Fermilab’s data and security requirements, all operations—including embedding generation and LLM inference—are performed within the DUNE internal computing environment. Only authenticated DUNE collaborators will be able to use this tool. We are exploring options to host the interface at Fermilab to provide secure, collaboration-wide access.

\section{Preliminary Results}
\begin{figure}[h!]
    \centering
    \includegraphics[width=0.8\textwidth]{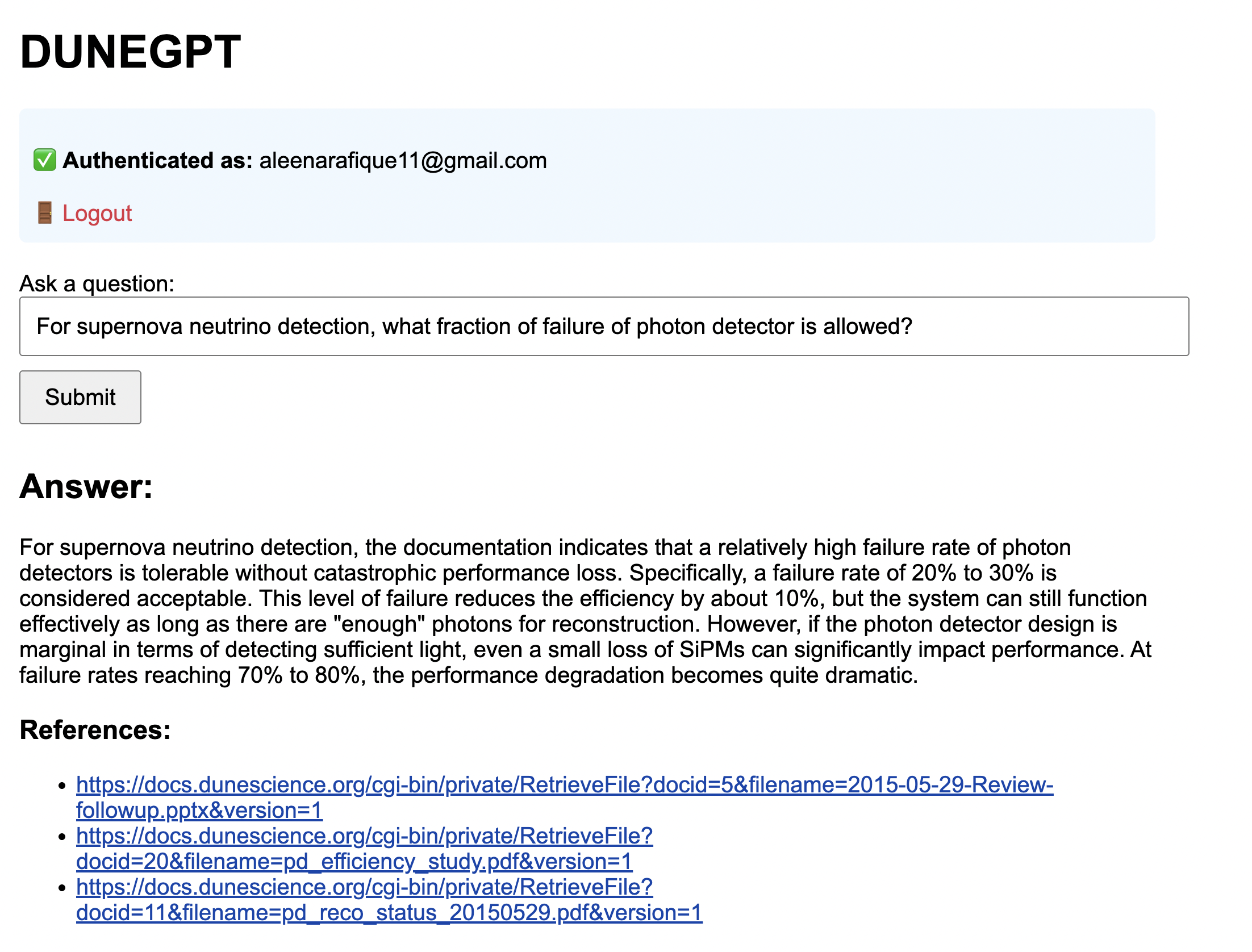}
    \caption{Frontend web interface showing a sample question, response, and the top three retrieved references used in response generation.}
    \label{fig:DUNE-GPT}
\end{figure}

Initial benchmarks demonstrate that the RAG-based system retrieves relevant documentation with high accuracy ($\sim70\%$) across diverse query types, including detector specifics, reconstruction algorithms, and physics analysis workflows. The tool is still under active development and will be opened for collaboration-wide testing soon. Figure~\ref{fig:DUNE-GPT} shows the current frontend and web interface with an example query and generated response, including document references.

Further improvements are planned to expand the embedding corpus, include multi-modal content (e.g., plots and figures), and integrate with version-controlled repositories. We are also working toward establishing web hosting at Fermilab for the frontend interface.

\section{Future Work}
Future development will focus on:
\begin{itemize}
    \item Expanding coverage to include detector operation logs and code documentation.
    \item Developing a RAG-based, domain-adapted LLM for HEP-specific language understanding.
    \item Deploying the tool for collaboration-wide use.
    \item Incorporating user feedback to improve retrieval quality and interface usability.
\end{itemize}

\section{Acknowledgments}
%
This document was prepared by DUNE collaboration using the resources of the Fermi National Accelerator Laboratory (Fermilab), a U.S. Department of Energy, Office of Science, Office of High Energy Physics HEP User Facility. Fermilab is managed by Fermi Forward Discovery Group, LLC, acting under Contract No. 89243024CSC000002.
%
%
This work was supported by
CNPq,
FAPERJ,
FAPEG, 
FAPESP and,
Funda\c{c}\~{a}o Arauc\'{a}ria,   Brazil;
CFI, 
IPP and 
NSERC,                          Canada;
CERN;
ANID-FONDECYT,                  Chile;
M\v{S}MT,                       Czech Republic;
ERDF, FSE+,
Horizon Europe, 
MSCA and NextGenerationEU,      European Union;
CNRS/IN2P3 and
CEA,                            France;
PRISMA+,                        Germany;
INFN,                           Italy;
FCT,                            Portugal;
CERN-RO/CDI,                        Romania;
NRF,                            South Korea;
Generalitat Valenciana, 
Junta de Andaluc\'{\i}a,
MICINN, and 
Xunta de Galicia,               Spain;
SERI and 
SNSF,                           Switzerland;
T\"UB\.ITAK,                    Turkey;
The Royal Society and 
UKRI/STFC,                      United Kingdom;
DOE and 
NSF,                            United States of America.

\end{document}